\documentclass[12pt]{spieman}  % SPIE journals manuscript class (incl. Advanced Photonics)

\usepackage{amsmath,amsfonts,amssymb}
\usepackage{graphicx}

\usepackage{setspace}
\usepackage{booktabs}
\usepackage{siunitx}

\newcommand{\chem}[1]{\ensuremath{\mathrm{#1}}}

\title{GiBS: Generative input-side basis-driven structures}

\author[a]{M Reza Marzban}
\author[a]{Ashkan Zandi}
\author[a,*]{Ali Adibi}

\affil[a]{School of Electrical and Computer Engineering, Georgia Institute of Technology, Atlanta, GA, USA}

\begin{document}
\maketitle

\begin{abstract}
Designing large-scale metasurfaces with nonlocal optical effects remains challenging due to the immense dimensionality and fabrication constraints of conventional optimization methods. We introduce \textbf{GiBS} (Generative Input-side Basis-driven Structures), an inverse-design framework that represents the entire device using a compact set of coefficients from smooth parametric bases such as Fourier or Chebyshev functions. This formulation compresses the design space by more than an order of magnitude, enabling efficient optimization of complex, broadband, and aperiodic geometries. GiBS integrates this low-dimensional representation with an autoencoder-based manifold-learning workflow to map the relationship between geometry and optical response, facilitating rapid exploration, discovery of high-performance designs, and systematic analysis of fabrication sensitivity. The inherent smoothness of the basis functions ensures manufacturability while capturing the asymmetry required for nonlocal optical interactions. We experimentally validated the framework through the realization of a PEDOT:PSS broadband scattering metasurface, whose measured response closely matched full-wave simulations across 500--1100\,nm. These results establish GiBS as a scalable, data-efficient, and fabrication-aware platform for the inverse design of multifunctional metasurfaces, bridging AI-guided representation learning with experimentally realizable photonic architectures.
\end{abstract}

% --- Keywords required (3--6 is typical)

% --- Corresponding author line (SPIE style)
{\noindent \footnotesize\textbf{*}Ali Adibi, \linkable{ali.adibi@ece.gatech.edu}}

% --- Double spacing for the manuscript body (SPIE template convention)
\begin{spacing}{2}

\section{Introduction}

The demand for compact, lightweight, and multifunctional optical components has driven advances in subwavelength control of electromagnetic waves. Metasurfaces, planar arrays of engineered nanostructures, provide an alternative to bulk optics and have enabled beam steering~\cite{abdollahramezani2022reconfigurable,wu2019dynamic}, lensing and imaging~\cite{li2022inverse,lin2021end}, optical neural networks~\cite{fu2024optical,zarei2020integrated,poordashtban2023integrated}, and polarization control~\cite{yu2014flat,kildishev2013planar}. Designing large-area or multifunctional devices, however, remains difficult. The parameter space is immense and non-convex~\cite{khaireh2023newcomer}; full-wave simulations are costly in time and memory~\cite{molesky2018inverse}, and; critically, the search itself suffers from the curse of dimensionality: even with a fast surrogate model to relate the design parameters to the output response. As a result, exhaustive or dense exploration of a high-dimensional design space remains infeasible.

A common strategy builds a dictionary of precomputed meta-atoms that are tiled to approximate desired phase or amplitude profiles~\cite{yu2014flat,chizari2016analog,abdollahramezani2020meta}. This relies on a local periodicity assumption that decouples neighboring responses. The assumption degrades for large deflection angles or strong nonlocal interactions~\cite{spagele2021multifunctional}, where collective lattice effects and high-Q (Q: quality factor) resonances~\cite{koshelev2018asymmetric}, such as quasi-bound states in the continuum (q-BICs)~\cite{hoang2025collective,hsu2016bound,overvig2021wavefront}, dominate. Resonant wavelengths can be sensitive to local phase gradients~\cite{malek2022multifunctional}, which favors smooth, continuous geometry variations that are not naturally produced by discrete dictionaries~\cite{spagele2021multifunctional}. Modern applications further demand multifunctionality, such as distinct behaviors across wavelengths, incident angles, or material states in phase-change systems~\cite{hemmatyar2021enhanced,marzban2025inverse_1}. This places additional strain on discrete catalogs and simple geometries. The combinatorial growth is severe: even a metasurface formed by a  $16\times16$ array of pixels with (coarse) 10 geometric choices (or design parameters) per pixel yields $10^{256}$ configurations~\cite{kiarashinejad2019deep}, illustrating the fundamental dimensionality bottleneck.

Among existing design approaches, gradient-based topological optimization (TO) can generate rich freeform layouts~\cite{jensen2011topology}, yet it often converges to local optima and requires many random restarts~\cite{christiansen2021inverse,marzban2025hilab} to yield a good (but nonoptimal) response. Multifunctional objectives introduce conflicting gradients, and aggregate metrics, such as worst-case or averaged figures of merit, can trap the optimizer in suboptimal regions~\cite{marzban2025hilab}. In addition, combining freeform two-dimensional geometrical features with physical parameters like height or period in TO is challenging. To improve fabrication tolerance, several studies compute three gradients per iteration—nominal, dilation, and erosion—and aggregate them~\cite{sell2019adjoint,wang2011projection}, or introduce projection and smoothing filters to enforce minimum feature sizes and binarization~\cite{hammond2025unifying,probst2024fabrication}. These steps, while essential for realistic fabrication, increase computational cost and can reduce nominal efficiency if not co-designed with the performance objective. Without them, the realized devices often deviate from simulated performance due to process variations. Alternatively, purely training-based inverse design models~\cite{kiarashinejad2020deep,liu2018generative,an2019deep,kiarashinejad2020knowledge} using artificial-intelligence algorithms face a different challenge: generating high-quality training datasets is expensive. Naive sampling produces numerous low-performing structures, and the curse of dimensionality persists even when predictions are fast~\cite{marzban2025hilab,marzban2025inverse}. Together, these issues motivate a framework that manages dimensionality, incorporates fabrication constraints naturally, and remains data-efficient.

To address these challenges, we look beyond the established geometric parameterizations used in inverse design. These include 1) pixel-based representations, where a device is described on a fine grid with tailored permittivity ~\cite{piggott2015inverse,jensen2011topology}, 2) level-set methods, which define layouts as contours of a scalar function~\cite{osher1988fronts,wang2003level}, 3) explicit shape parameterizations, where analytic functions map a few variables to constrained geometries~\cite{chen2020design,zhou2024inverse,gershnabel2022reparameterization}, and 4) neural parameterizations, where layouts are represented by analytic neural networks~\cite{dai2025shaping}. Within the broader lens of \textit{input-side representation learning}~\cite{marzban2025inverse}, we report here an effective fifth class: a global, basis-driven representation. \textbf{GiBS} (\textbf{G}enerative \textbf{I}nput-side \textbf{B}asis-driven \textbf{S}tructures) describes the entire device as a continuous function defined by a small set of coefficients in a chosen parametric basis, such as Fourier or Chebyshev series~\cite{moharam1981rigorous,agrawal2012fiber}. The key aspect of this approach is the assumption that optimal structures do not vary dramatically with space. This is observed in many reported mestasurfaces for practical applications. The resulting spatial band-limited nature of the permittivity profile can be used to represent the design parameters in terms of a set of bases to considerably reduce the dimensionality of the problem. This smooth, low-dimensional representation directly addresses the need for continuous geometry variations in nonlocal metasurfaces, transforming a combinatorial search into an optimization over a compact, physically interpretable parameter space. Recent findings suggest that even complex freeform devices can be represented by a remarkably small number of latent variables~\cite{marzban2025hilab}, supporting the efficiency and interpretability of the basis-driven approach.

Manufacturability is incorporated in GiBS at the representation level. Instead of defining arbitrary freeform boundaries, GiBS modulates the parameters of a robust family of primitive shapes, such as nanopillar radii, that are naturally compatible with standard lithographic and etching processes. Simple, convex features are fabricated with higher fidelity and yield~\cite{khorasaninejad2016metalenses,arbabi2015subwavelength}, whereas complex freeform contours often exhibit reduced reliability~\cite{schubert2022inverse,pestourie2018inverse}. While freeform pipelines can include fabrication-aware steps such as subpixel smoothing, projection filters, and minimum-feature constraints, or rely on foundry-compatible workflows~\cite{hammond2025unifying,pan2023fabrication,kim2024lowloss,chen2025inverse}, these constraints are typically introduced post hoc. GiBS embeds manufacturability by construction through its basis vocabulary, ensuring fabrication-tolerant designs without additional penalty terms~\cite{yu2014flat,he2018high,chen2025machine}.

This representation also mitigates the tension between device scale and computational feasibility. Direct pixel optimization of large supercells is intractable~\cite{li2022inverse,kang2024large}, whereas local-periodicity approximations neglect long-range coupling~\cite{zhou2024inverse,chen2024inverse}. By operating with a substantially reduced number of degrees of freedom, GiBS enables the practical optimization of broadband and worst-case optical responses that would be computationally prohibitive in pixel-based formulations ~\cite{ahadi2025bayesian}. In GiBS, low-order coefficients capture global correlations across the aperture, and higher-order coefficients refine subwavelength details~\cite{bi2025spatial}, maintaining tractability while preserving physical continuity. This compact description supports multiple optimization pathways. For instance, manifold learning~\cite{zandehshahvar2022manifold} can analyze the latent response manifolds generated by different bases, guiding basis selection. The same coefficients can be used in gradient-based shape optimization~\cite{gershnabel2022reparameterization,thelen2022multi}, enabling direct $\partial \text{FoM}/\partial A_k$ (FOM: Figure of Merit, $A_k$: the $k$-th basis)
updates in a relaxed space with a handful of design variables rather than hundreds of pixels. This not only simplifies the optimization but also enforces manufacturability through its primitive basis functions. Furthermore, fabrication robustness can be assessed by introducing Gaussian perturbations on the coefficients and selecting designs with minimal response deviation. The remainder of this paper presents the theoretical and computational framework of GiBS and concludes with its experimental validation through the design of a wideband transparent beam scatterer.

\section{A Parametric Basis Framework for Nonlocal Metasurfaces}

\subsection{Dimensionality Reduction with Basis Expansions}
\label{sec:basis}

The core of the GiBS framework is the parameterization of the metasurface geometry using a basis expansion. To make the geometric construction explicit, we first define the metasurface supercell as a square array of cylindrical nanopillars whose local radii are prescribed by the continuous function $R(x,y)$. This can be mathematically formulated as Eq.~\eqref{eq:basis_general}, which assigns a radius to each discrete pillar at location $(x,y)$ by evaluating the continuous function $R(x,y)$ over the supercell domain. The radius is expressed as a linear combination of orthogonal basis functions $\phi_{n_x,n_y}(x,y)$, governed by a compact set of coefficients $A_{n_x,n_y}$ that define the overall device morphology:
\begin{equation}
    R(x,y) = F\left( \sum_{n_x=0}^{N_x} \sum_{n_y=0}^{N_y} A_{n_x,n_y} \phi_{n_x,n_y}(x,y) \right)
    \label{eq:basis_general}
\end{equation}
\begin{figure}[h!]
    \centering
    \includegraphics[width=1\textwidth]{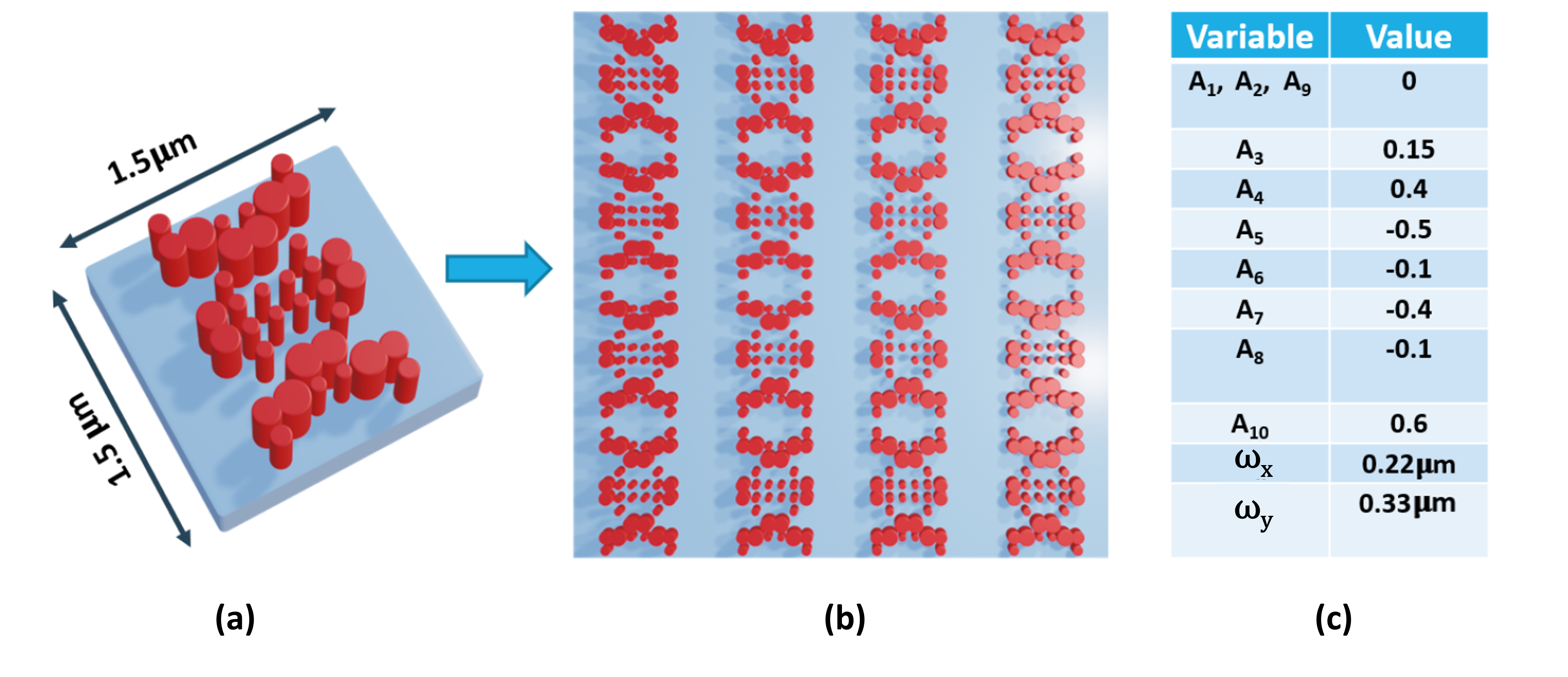}
    \caption{%
    Parametric basis modeling in the GiBS framework. (a) A single supercell geometry generated using Eq.~\eqref{eq:basis_general} and~\eqref{eq:basis_sinusoidal} with coefficients in (c), showing a smooth and asymmetric variation of pillar radii. (b) A periodic metasurface formed by the pixel in (a), demonstrating how continuous control of the basis terms produces globally varying yet fabrication-friendly layouts. (c) Representative values of the basis coef-ficients $A_k$ and frequency parameters ($\omega_x$, $\omega_y$) defining the nanopillar distribution.}
    \label{fig:basis_modeling}
\end{figure}
The representative unit-cell example illustrated in Figure~\ref{fig:basis_modeling}(a) is generated on a $16\times16$ grid (256 candidate pillar sites in one unit cell that creates the large periodic structure in Figure~\ref{fig:basis_modeling})(b), where $R(x,y)$ is fully determined by 12 basis coefficients, together with two additional parameters specifying the supercell period and the pillar thickness. The particular realization shown in Figure~\ref{fig:basis_modeling}(a) uses the Fourier expansion in Eq.~\eqref{eq:basis_sinusoidal}, with amplitude coefficients $A_k$ taken directly from the values listed in Figure~\ref{fig:basis_modeling}(c).
\begin{equation}
    R(x,y) = r_{\text{base}} \cdot \left( A_0 + \sum_{k=1}^{N} \left[ A_{2k-1}\sin\!\left(k\omega_x \frac{2\pi x}{P_x} + k\omega_y \frac{2\pi y}{P_y}\right) + A_{2k}\cos\!\left(k\omega_x \frac{2\pi x}{P_x} + k\omega_y \frac{2\pi y}{P_y}\right) \right] \right),
    \label{eq:basis_sinusoidal}
\end{equation}
where $r_{\text{base}}$ is the base radius, $A_k$ are amplitude coefficients, $\omega_x$ and $\omega_y$ are spatial frequency parameters, and $P_x$ and $P_y$ are the supercell periodicities. The inherent smoothness of Fourier modes ensures gentle spatial variations, minimizing abrupt phase transitions and promoting stable optical coupling across the unit cell (or the aperture). Although $R(x,y)$ is defined on all 256 grid points, only the pillars whose radii exceed the fabrication threshold imposed by the nonlinear mapping in Section 2.2 appear in Figure~\ref{fig:basis_modeling}(a), yielding the observed 36-pillar configuration. Overall, Figure~\ref{fig:basis_modeling} shows how a small set of parameters can encode complex and smoothly varying designs. Here, the coefficients $A_{n_x,n_y}$ serve as the reduced design variables to be optimized, while $F(\cdot)$ denotes a nonlinear mapping that enforces fabrication constraints, as discussed in Section~\ref{sec:constraints}. 
The choice of basis functions $\phi_{n_x,n_y}$ acts as a structural prior, guiding the model toward geometries that naturally satisfy smoothness and periodicity requirements. In this work, the basis functions control the local radius of cylindrical nanopillars at each point $(x,y)$; however, the same formalism can be extended to other unit-cell geometries such as hexagonal posts, rectangular blocks, or even freeform scatterers. Cylindrical pillars were selected in this study because their smooth, edge-free profiles simplify fabrication and mitigate lithographic and etching-induced errors, providing a consistent test platform for validating the framework. These pillars are also common choices in a large variety of reported structures for several applications~\cite{li2022inverse,abdollahramezani2021dynamic,kiarashinejad2020deep,kuznetsov2024roadmap}. 

This formulation draws a useful analogy to rigorous coupled-wave analysis (RCWA), where a known
permittivity distribution is expanded into a Fourier series to solve Maxwell’s equations~\cite{moharam1995stable}.
In GiBS, we invert this logic: the basis coefficients themselves become the primary design variables that generate
the structure, forming a bridge between the mathematical language of simulation and the optimization process.
To illustrate this concept, we employ two representative basis families, Fourier and Chebyshev expansions, which exemplify periodic and finite-aperture configurations, respectively, and collectively demonstrate the
generality of the GiBS formulation.

\paragraph{Fourier Basis.}
A two-dimensional Fourier series provides a natural framework for periodic structures that form the foundation of many metasurfaces. The sinusoidal expansion is expressed as Eq.~\eqref{eq:basis_sinusoidal}.

\paragraph{Chebyshev Basis.}
For finite-aperture or aperiodic devices, Chebyshev polynomials provide an effective alternative. The Chebyshev polynomials of the first kind, $T_n(u)$, are defined on the interval $[-1, 1]$ by the recurrence relation $T_{n+1}(u)=2uT_n(u)-T_{n-1}(u)$, with $T_0(u)=1$ and $T_1(u)=u$~\cite{mason2002chebyshev}. A two-dimensional expansion can then be constructed as
\begin{equation}
    \phi_{n_x,n_y}(x,y) = T_{n_x}\!\left(\frac{2x - (x_{\text{max}} + x_{\text{min}})}{x_{\text{max}} - x_{\text{min}}}\right) 
    T_{n_y}\!\left(\frac{2y - (y_{\text{max}} + y_{\text{min}})}{y_{\text{max}} - y_{\text{min}}}\right),
    \label{eq:chebyshev}
\end{equation}
where the arguments are scaled to map the rectangular domain $[x_{\text{min}}, x_{\text{max}}]\times[y_{\text{min}}, y_{\text{max}}]$ to the canonical square $[-1,1]^2$. Unlike periodic Fourier series, Chebyshev polynomials emphasize variations near the boundaries, which is advantageous for controlling edge effects and boundary-induced behavior in finite-aperture metasurfaces. 

In this work, both Fourier and Chebyshev bases are tested to demonstrate the versatility of the GiBS framework; however, the approach is not limited to these two choices. Other basis families, such as Bessel, Legendre, or wavelet expansions, can be explored for specific applications where the underlying optical modes or physical constraints suggest more suitable functional forms. Selecting or tailoring the basis to match the dominant modal behavior of a given device represents a natural direction for future extensions of this framework.

\subsection{Enforcing Asymmetry and Fabrication Constraints}
\label{sec:constraints}

The basis-expansion framework inherently supports asymmetric geometries, which are essential for nonlocal metasurface functionalities such as beam splitting, asymmetric scattering, and polarization conversion. Asymmetry can be introduced naturally by including both sine and cosine components in the Fourier expansion or by incorporating odd-order Chebyshev terms. This enables fine control over spatial phase gradients and coupling pathways without increasing dimensionality.

To bridge the continuous mathematical formulation with the discrete domain of nanofabrication, the GiBS model employs a nonlinear mapping function $F(\cdot)$. This function enforces physical and manufacturing limits, ensuring that generated layouts remain fabrication-compatible. A simple yet effective choice is a threshold-based rule that suppresses geometrical features below a certain size while preserving the continuous structure of the rest of the design. This operation is particularly useful for lithographic patterning and binary layout generation. One representative form is
\begin{equation}
    F(\alpha) = 
    \begin{cases}
        r_{\text{low}}, & \text{if } \alpha < \theta, \\
        \alpha, & \text{otherwise,}
    \end{cases}
    \label{eq:binarization}
\end{equation}
where $\alpha$ is the continuous output of the basis expansion, and $\theta$ is a tunable threshold that defines the minimum printable feature. In this implementation, when the evaluated radius falls below the threshold, the corresponding pillar is removed (by setting $r_{\text{low}}=0$), or optionally replaced by a minimum viable radius to retain structural continuity. Values above the threshold remain unchanged, meaning that $R(x,y)$ directly follows the continuous basis expansion for fabricable regions. This selective truncation approach maintains smooth variations where possible while eliminating highly miniaturized features that cannot be reliably fabricated. By embedding such fabrication awareness directly into the generative model, the optimization explores only physically realizable regions of the design space. This improves convergence efficiency, prevents wasted iterations on non-manufacturable geometries, and enhances the likelihood of experimental success.

\subsection{Design Platform and Problem Setting for GiBS Evaluation }
To demonstrate and evaluate GiBS, we apply it to the inverse design of a broadband scattering metasurface in the conducting polymer PEDOT:PSS~\cite{fan2019pedot,doshi2024direct,ludescher2025femtosecond,lee2023dynamic}. PEDOT:PSS poses a demanding design space: its real refractive index is relatively low in the visible and increases toward the near-infrared, while phase-dependent extinction introduces additional loss in the metallic state~\cite{dingler2022situ}. Achieving efficient, wide-angle scattering, therefore, requires coordinating local resonances (set by nanopillar geometry) with nonlocal, supercell-scale coupling. GiBS addresses this challenge by combining low-dimensional geometric encoding with data-driven response modeling. In addition to its optical complexity, PEDOT:PSS is deliberately selected as the material platform for two reasons. First, its relatively low refractive index makes high-efficiency scattering considerably more difficult compared to high-index platforms such as TiO$_2$ or Si. Demonstrating broadband, wide-angle functionality on such a low-index material therefore, highlights the expressive power of the basis-driven GiBS parameterization. Second, PEDOT:PSS is an actively tunable material whose refractive index and absorption can transition between insulating and metallic states under applied electrical bias~\cite{dingler2022situ}. This tunability makes it a compelling platform for studying multifunctional metasurfaces, where identical geometries may exhibit qualitatively different optical responses across material phases~\cite{yin2017beam,karst2021electrically}. In this work, experimental validation is focused on the insulating phase, with PEDOT:PSS serving as a challenging testbed for assessing the effectiveness of GiBS as an inverse-design framework and motivating its extension to future multi-state and reconfigurable metasurface designs.

\subsection{Manifold Learning for Reducing the Dimensionality of Electromagnetic Responses }

Manifold learning is a powerful approach in reducing the dimensionality of the response space of electromagnetic structures~\cite{zandehshahvar2022manifold},  especially when combined with proper metrics~\cite{zandehshahvar2023metric}. Among different manifold-learning algorithms, autoencoder has been extensively studied for the design of metaphotonic structures~\cite{kiarashinejad2020deep,kiarashinejad2020knowledge}. Combining GiBS with manifold learning of the response space enables better representation and visualization of the response space and provides a measure for selecting the number of bases for a given problem, as well as comparing different options for the basis functions. In this work, we will use an autoencoder with a properly selected metric (or loss function) for dimensionality reduction. 

\begin{figure}[h!]
    \centering
    \includegraphics[width=\textwidth]{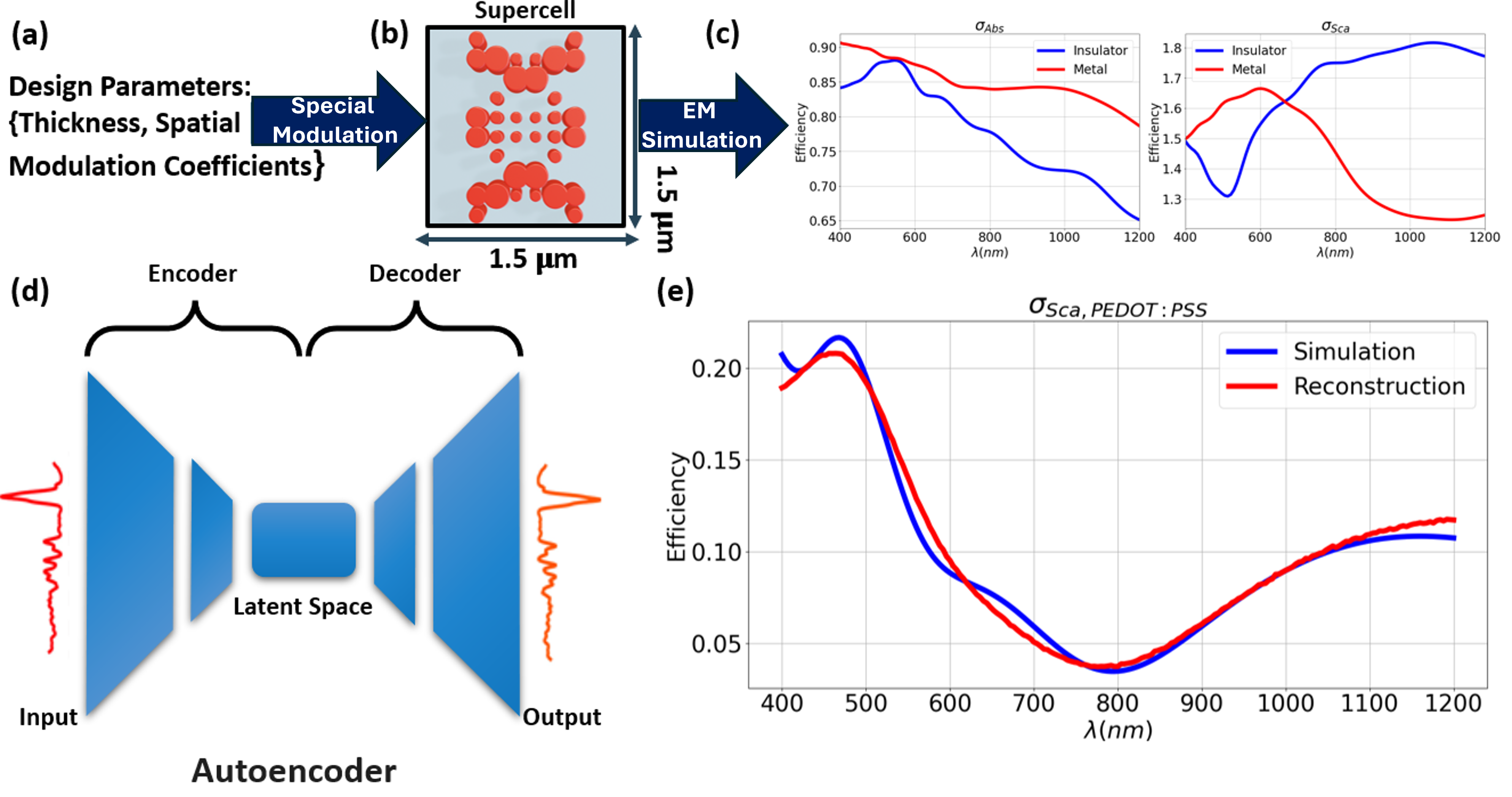}
    \caption{Workflow of the autoencoder-assisted design pipeline. (a) Design parameters (basis coefficients) define (b) a supercell geometry, which is simulated to yield (c) electromagnetic spectra, including both scattering and absorption cross-sections for different material states. (d) A trained autoencoders compresses each one of these spectral responses into a low-dimensional latent space, and (e) reconstructs them with high accuracy, validating the fidelity of the learned manifold. Four separate autoencoders were trained in total, one for each combination of response type (absorption, scattering) and material phase (insulating, metallic).}
    \label{fig:workflow}
\end{figure}

Figure~\ref{fig:workflow} depicts the flow of our analysis and design approach. After selecting a set of basis functions with a given number of coefficients (Figure~\ref{fig:workflow}(a)), e.g., through prior knowledge or by pure trial and error, to specify a super-cell Figure~\ref{fig:workflow}(b)), we form a series of structures by randomly selecting the basis coefficients and simulating the resulting structure using the three-dimensional finite-difference time-domain (3D FDTD) technique (Lumerical) over the 400--1200~nm wavelength range to obtain the response in terms of absorption and scattering cross sections ($\sigma_{\mathrm{abs}}(\lambda)$ and $\sigma_{\mathrm{sca}}(\lambda)$, respectively) in the two material phases of PEDOT:PSS, as shown in Figure~\ref{fig:workflow}(c). These two functions are uniformly sampled at 201 wavelengths in the 400--1200~nm range to form a 201-dimensional response vector. The resulting design--response pairs are used to train an autoencoder that compresses each 201-point spectral response into a compact latent representation (Figure~\ref{fig:workflow}(d)). The detailed information of the autoencoder (number of layers and nodes in each layer, especially the size of the bottleneck layer) is found by combining prior knowledge about such structures and trial-and-error. Once properly trained, reconstructions from the latent response space (i.e., through the decoder part of the autoencoder) closely match the simulated spectra (Figure~\ref{fig:workflow}(e)), indicating that the learned latent space retains the salient physics of the electromagnetic response.

\paragraph{Architecture and training.}
For each response type and material phase we train an independent, fully connected, symmetric autoencoder, yielding four models in total: $\{\sigma_{\text{abs}}, \sigma_{\text{sca}}\}\times\{$insulator, metal$\}$. Each network takes a 201-point spectrum sampled uniformly over wavelength as its input. The encoder maps $\mathbb{R}^{201}\!\to\!\mathbb{R}^2$ through the sequence 201\,→\,90\,→\,60\,→\,50\,→\,20\,→\,2 using $\tanh(\cdot)$ activations, and the decoder mirrors this mapping back to 201 with a linear output layer. The two-dimensional bottleneck enables visualization and clustering of responses while retaining sufficient representational capacity for smooth spectra reconstruction. The model is trained using a combined loss function that balances spectral similarity and amplitude accuracy,
\begin{equation}
\mathcal{L}(y,\hat{y})=\lambda_1\!\left[1-\frac{y\cdot \hat{y}}{\lVert y\rVert_2\,\lVert \hat{y}\rVert_2}\right]+\lambda_2\,\frac{1}{N}\lVert y-\hat{y}\rVert_1,
\qquad \lambda_1=0.7,\ \lambda_2=0.3,
\end{equation}
where $y$ and $\hat{y}$ are the true and reconstructed spectra and $N=201$. This combination of cosine and absolute losses enforces both phase and magnitude consistency. Each network is trained for 20 epochs on a dataset of 6,000 simulated spectra, with an 80/20 train–test split and validation monitoring. The training set includes three geometry families, random layouts, Fourier-basis GiBS designs, and Chebyshev-basis GiBS designs, ensuring the latent manifold captures both structured and unstructured response regimes. The final validation losses converge to the order of $10^{-4}$ on normalized spectra, consistent with the reconstructions shown in Figure~\ref{fig:workflow}e.

This combination of geometric compression (input-side representation) and spectral compression (output-side representation) establishes a closed-loop framework where both the structure and its optical response are embedded in tractable, continuous manifolds. The GiBS representation thus provides the foundation for systematic exploration of nonlocal metasurface behaviors in both physical and functional domains.

\subsection{Latent Space Analysis for Inverse Design and Active Optics}

\begin{figure}[h!]
    \centering
    \includegraphics[width=\textwidth]{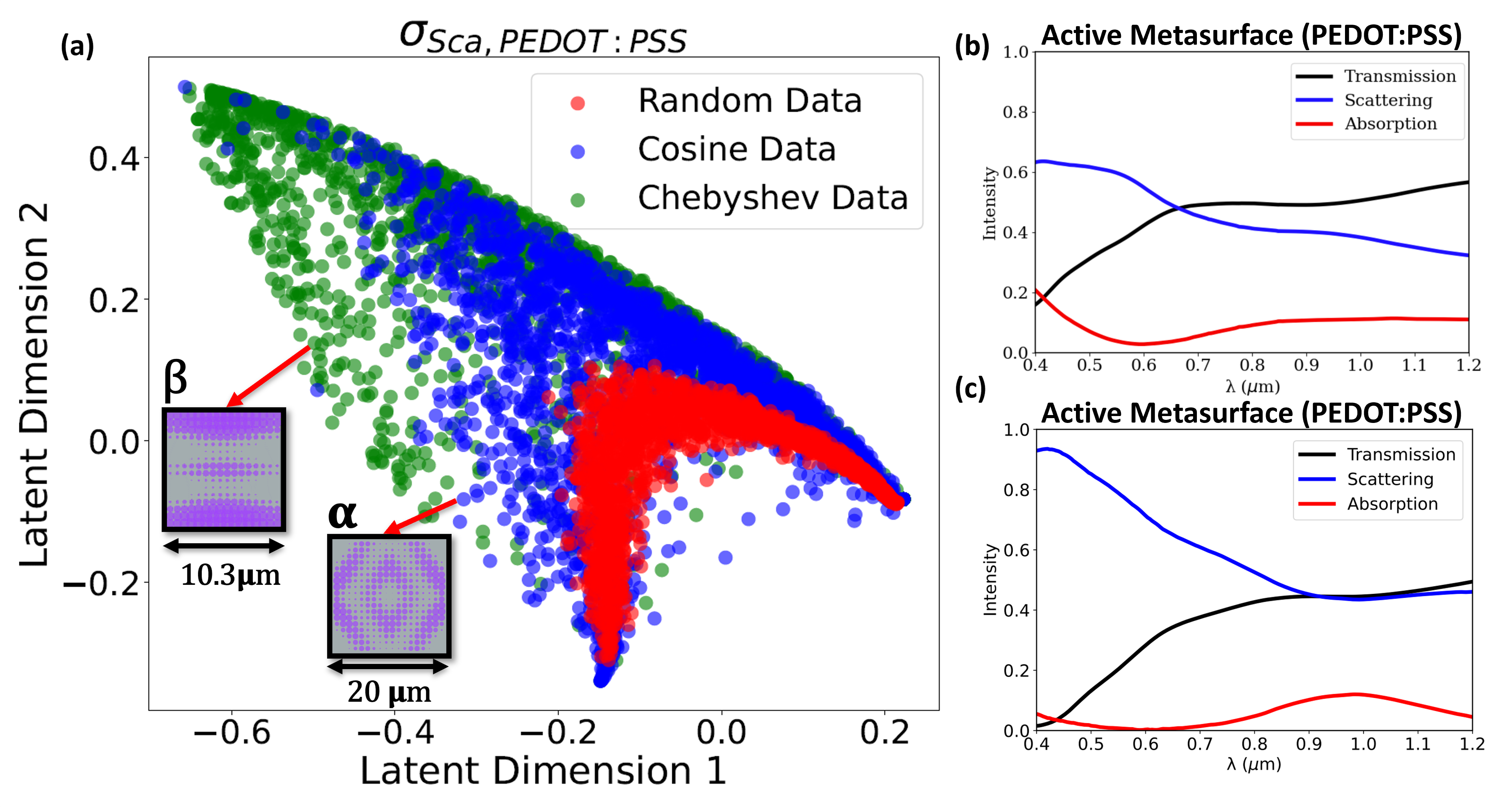}
    \caption{Latent-space analysis of the GiBS framework applied to PEDOT:PSS metasurfaces with structures in Figure~\ref{fig:basis_modeling} in the insulating phase. 
    (a) Two-dimensional latent embedding of the scattering cross-sections $\sigma_{\text{sca}}$ response showing the distribution of responses from random (red), Fourier cosine (blue), and Chebyshev (green) basis-generated geometries. The Fourier and Chebyshev parameterizations span a broader and more continuous manifold compared to random designs. Insets $\alpha$ and $\beta$ mark representative 400\,nm-thick PEDOT:PSS metasurfaces on SiO$_2$ substrates with supercell periods of 20\,$\mu$m and 10.3\,$\mu$m, respectively. 
    (b) Optical response of the Fourier (cosine) basis design $\alpha$, showing uniform scattering and transmission balance across the visible and near-infrared spectrum. 
    (c) Optical response of the Chebyshev-based design $\beta$, exhibiting enhanced scattering intensity in the visible range due to strong boundary modulation. 
    These examples illustrate how different basis functions can generate distinct yet physically consistent electromagnetic responses within the unified GiBS design manifold.}
    \label{fig:latent_exploration}
\end{figure}

The learned latent space acts as a navigable map linking geometry to functionality. This organization enables an intuitive inverse design workflow: target optical responses can be associated with regions of the latent manifold, from which new candidate geometries are decoded via their basis coefficients. With each point in the latent space corresponding to a distinct response, the range of the latent space covered by the responses from a structure is an indication of the ability of that structure in forming a diverse set of responses (i.e., design flexibility). This allows us to compare different classes of structures and/or different modeling techniques.A key outcome is that the GiBS framework, through its structured basis encoding, substantially enlarges the accessible region of the performance landscape relative to random sampling.. For this investigation, we simulated a total of 12000 structures with 4000 generated using each class: random, Fourier, and Chebyshev bases. In our dataset, the PEDOT:PSS thickness is fixed at 400 nm, while the supercell period (1–40 µm), the number of pillars, and the basis coefficients are varied. As shown in Figure~\ref{fig:latent_exploration}a, structures in Figure~\ref{fig:basis_modeling} generated by random nanopillar dimensions (red) occupy a limited region of the latent space, whereas those constructed through Fourier (cosine-only, blue) or Chebyshev (green) basis expansions cover a much broader and continuous manifold, corresponding to a richer diversity of spectral behaviors. The two basis families exhibit complementary properties: Fourier bases produce smooth, periodic variations suited to diffractive or uniformly scattering systems, while Chebyshev bases localize variations near boundaries, enhancing finite-aperture effects and boosting overall scattering strength in the visible region. It is also clear that Chebyshev’s bases allow a wider range of responses (i.e., a larger area in the latent space covered by green dots compared to the dot with other two colors). 
Figures~\ref{fig:latent_exploration}b and~\ref{fig:latent_exploration}c illustrate two representative broadband scatterers selected from high-performing latent regions marked as $\alpha$ and $\beta$ in Figure~\ref{fig:latent_exploration}a. The Fourier (cosine) basis design $\alpha$ exhibits a uniform and balanced scattering response across 400–700\,nm, whereas the Chebyshev-based design $\beta$ produces stronger overall scattering intensity in the visible range (and weaker in the near-infrared). These results demonstrate that different functional bases can yield distinct optical characteristics highlighting the versatility of GiBS in tailoring metasurface behavior through basis selection. Combining the designs found using the two bases will enhance the range of available responses from a given structure. 
This dual perspective, input-side geometric control via basis coefficients and output-side response analysis via manifold learning, captures the central concept of GiBS. The input representation enforces continuity, manufacturability, and spatially uniform sensitivity to errors, while the learned output manifold organizes the high-dimensional optical response space into interpretable clusters. Together, these elements allow systematic exploration of how geometry, material composition, and tunable parameters govern performance. 
\begin{figure}[h!]
    \centering
    \includegraphics[width=\textwidth]{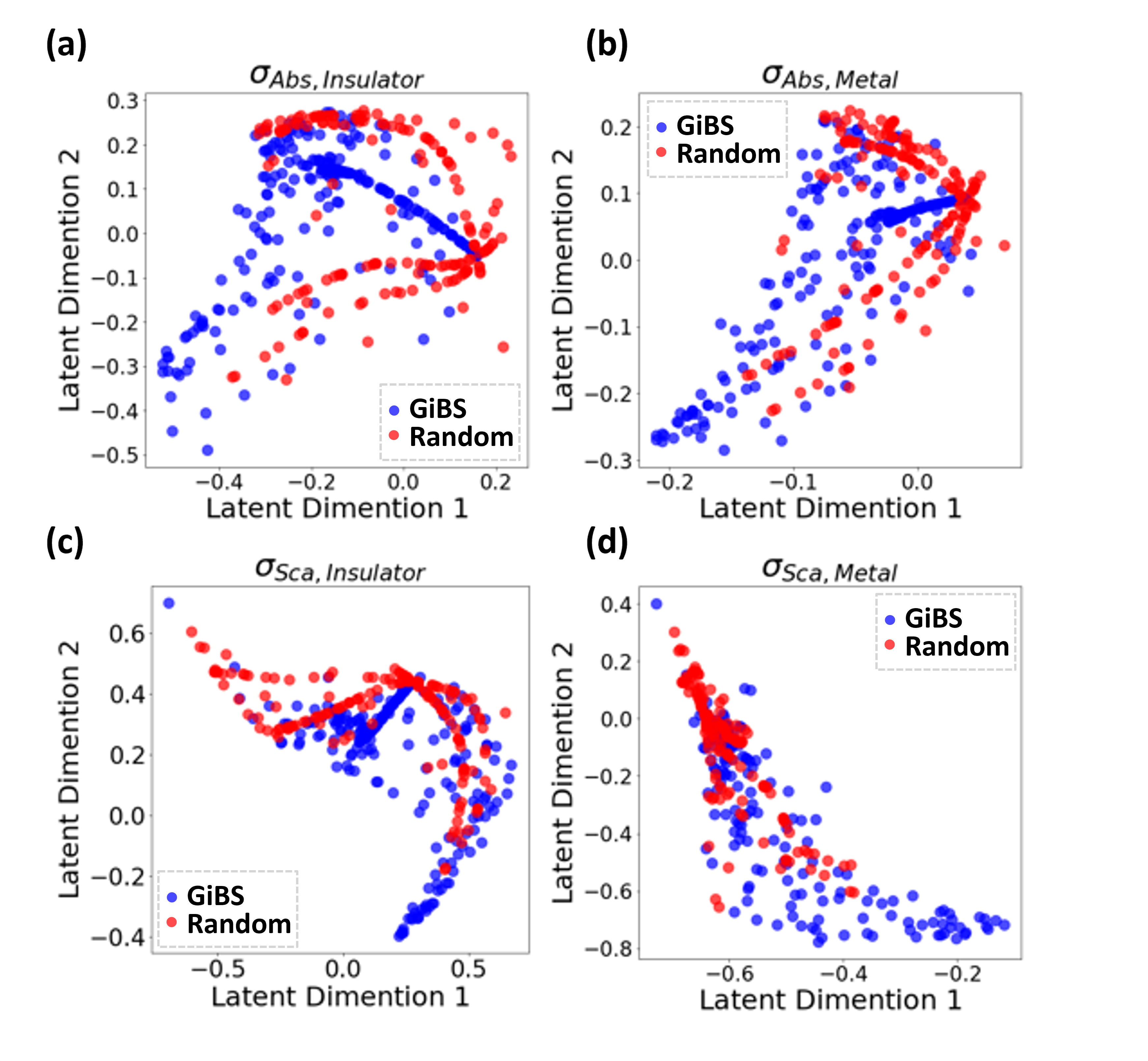}
    \caption{
    Latent-space distributions of $\sigma_{\text{abs}}$ and $\sigma_{\text{sca}}$ spectra for PEDOT:PSS metasurface geometries, in which 
    (a) $\sigma_{\text{abs}}$ in the insulating state, 
    (b) $\sigma_{\text{abs}}$ in the metallic state, 
    (c) $\sigma_{\text{sca}}$ in the insulating state, and 
    (d) $\sigma_{\text{sca}}$ in the metallic state 
    are visualized for GiBS-generated (blue) and random (red) geometries. 
    Corresponding embeddings for GiBS-generated (Fourier basis) designs show that the structured parameterization broadens and smooths the latent coverage, revealing continuous and connected manifolds across both insulating and metallic states.
    }
    \label{fig:active_gibs}
\end{figure}
While the previous section focused on designing broadband scattering in the insulating phase of PEDOT:PSS, the GiBS framework also provides a powerful platform for analyzing and optimizing \emph{active metasurfaces}, systems that exhibit multiple optical states with distinct refractive indices. To illustrate this capability, we consider a 400\,nm-thick PEDOT:PSS film on a thick silicon oxide (SiO$_2$) substrate as a representative active material platform, modeled in both its insulating and metallic phases, and seek to expand the optical response space across these states. The objective here is to simultaneously capture and expand the diversity of responses across both phases, enabling inverse design of metasurfaces that can operate under dual-state objectives. For instance, one may target a design that functions as a strong broadband scatterer in the insulating phase while acting as an efficient broadband absorber in the metallic phase, a configuration that is both spectrally broadband and functionally reconfigurable. As shown in Figure~\ref{fig:active_gibs}, the GiBS framework exhibits a consistent trend of expanding the accessible response space across material states. In this study, the design parameters include both the basis coefficients and the PEDOT:PSS layer thickness, with separate autoencoders trained for the insulating and metallic phases.
To ensure a fair and controlled comparison between random geometries and the GiBS (Fourier-basis) representation, we generated 2,000 designs for each case, randomly sampled geometries and GiBS-parameterized ones, and simulated both material states for every design. This equal sampling of the design space is essential for quantifying the true impact of the basis-driven representation on coverage, continuity, and the diversity of achievable optical responses. As shown in Figure \ref{fig:active_gibs}, the GiBS representation significantly expands and organizes the accessible response space across both phases. Random geometries occupy narrow, fragmented latent regions with limited continuity, indicating that brute-force sampling with a finite number of simulations fails to uncover the full range of tunable behaviors. In contrast, the Fourier-basis representation produces broader, smoother, and more connected manifolds across all absorption and scattering modalities. This structured input-side representation is particularly advantageous in multi-state design settings, since it ensures that responses from both the insulating and metallic phases populate a common, well-organized latent space. This expansion of the latent coverage demonstrates that GiBS not only improves the design diversity within a single phase but also provides a unified foundation for dual-state optimization, enabling devices with complementary functionalities across different material conditions. Once these compact, well-organized manifolds are established, gradient-based or Bayesian optimization methods can be applied to identify Pareto-optimal designs that balance conflicting objectives, such as maximizing scattering in one material state while enhancing absorption in the other.

\section{Experimental Validation with a PEDOT:PSS Metasurface}
To experimentally validate the full GiBS design-to-fabrication workflow, a large-area PEDOT:PSS metasurface was realized on a fused silica (\chem{SiO_2}) substrate. The optimized design, generated through the GiBS framework, corresponds to a \SI{25}{\micro\meter}~$\times$~\SI{25}{\micro\meter} supercell containing nanopillars with diameters ranging from \SI{80}{\nano\meter} to \SI{900}{\nano\meter}. This parametric distribution enables the coexistence of both local and nonlocal resonances within a single supercell, an essential mechanism for achieving broadband scattering spanning from 500 to 1100\,nm, with the spectral range bounded by the spectrometer in the characterization setup. 
\begin{figure}[h!]
    \centering
    \includegraphics[width=\textwidth]{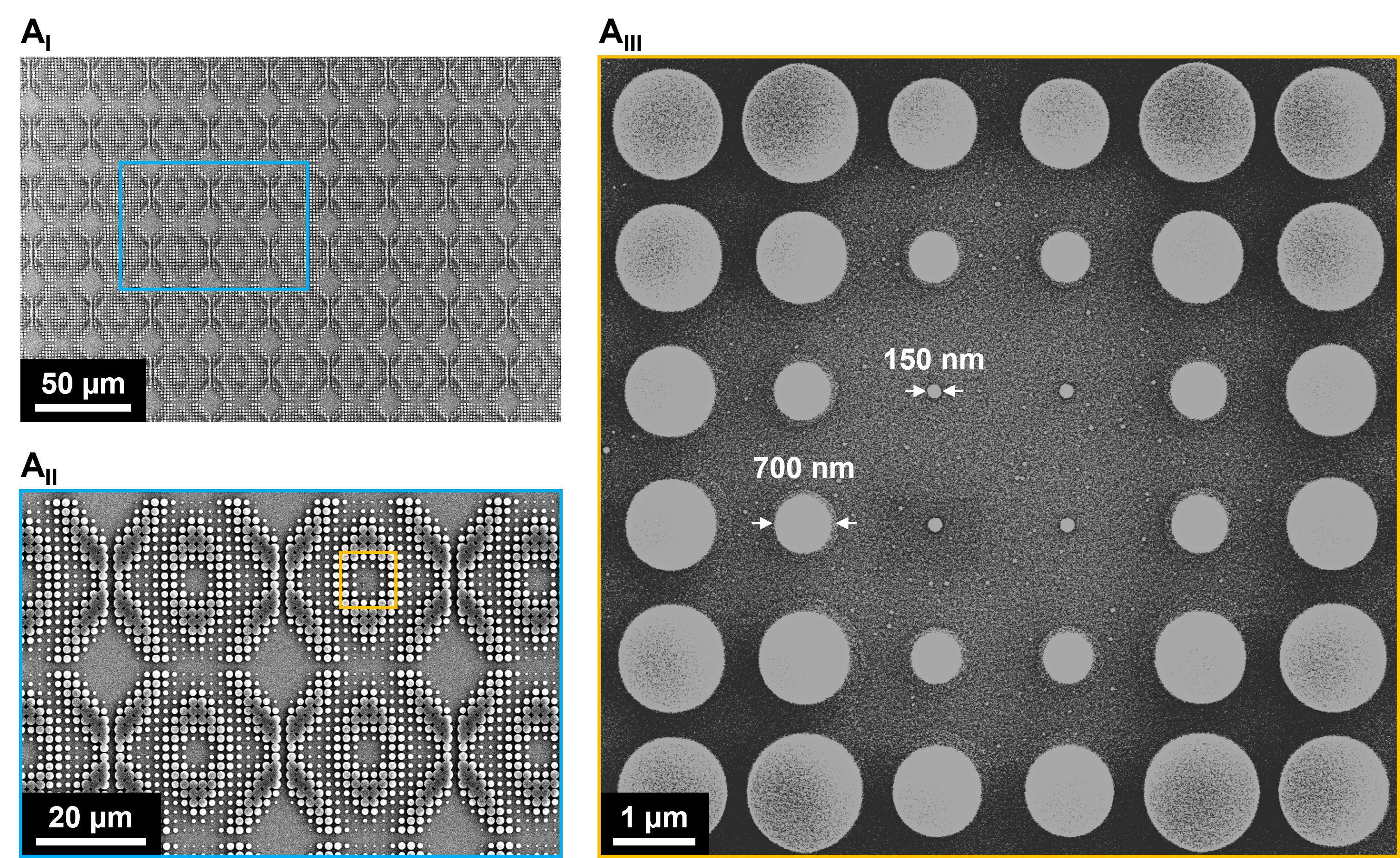}
    \caption{SEM images of the fabricated PEDOT:PSS metasurface optimized via the GiBS framework. (A$_\mathrm{I}$) Large-area view showing the periodic tiling of the 25\,$\mu$m\,$\times$\,25\,$\mu$m supercell. (A$_\mathrm{II}$) Magnified view of the region outlined in blue, showing the basis-driven nanopillar arrangement within individual supercells and its close agreement with the designed geometry. (A$_\mathrm{III}$) High-magnification view of the region outlined in orange near the supercell center, where the smallest and largest pillars sit side by side. The marked pillars measure 150\,nm and 700\,nm in diameter. The pillars span designed diameters from 80\,nm to 900\,nm in a 400\,nm-thick PEDOT:PSS film. Having both sub- and superwavelength features in the same supercell supports local and nonlocal resonances, which enables broadband scattering. The pillars have well-defined edges at every length scale shown, which indicates high pattern-transfer fidelity.}
    \label{fig:sem_pedot}
\end{figure}
\subsection{Fabrication Process}
The fabrication process followed a multi-step protocol to ensure high fidelity and reproducibility. The \chem{SiO_2} substrate was first cleaned by sequential rinsing in acetone and isopropyl alcohol (IPA), followed by an oxygen plasma treatment to enhance surface wettability. The PEDOT:PSS film was then deposited using a two-step spin-coating procedure with Heraeus Clevios PH1000 solution, filtered and mixed with IPA to improve uniformity. Each coating was spun at \SI{3000}{rpm} for \SI{60}{s} and baked at \SI{120}{\celsius} for \SI{15}{min}, producing a total film thickness of \SI{400}{\nano\meter}. For patterning, a Polymethyl methacrylate (PMMA) A4 resist layer was spin-coated and baked prior to electron-beam (e-beam) lithography (EBL). The optimized GDS file, generated directly from the GiBS parameterization, was written into the resist and developed in a 1:3 mixture of MIBK:IPA (MIBK: Methyl isobutyl ketone). A \SI{30}{\nano\meter} \chem{SiO_2} hard mask was then deposited by e-beam evaporation, followed by lift-off in acetone. The pattern was transferred into the PEDOT:PSS layer using argon dry etching, after which the residual hard mask could be optionally removed using buffered oxide etchant (BOE). Figure~\ref{fig:sem_pedot} shows scanning electron microscopy (SEM) images of the fabricated structure. The device demonstrates excellent pattern-transfer fidelity, with smooth transitions between adjacent nanopillars and uniform periodicity across the full area. The smallest features (\SI{80}{\nano\meter}) and the largest pillars (\SI{900}{\nano\meter}) are clearly resolved, confirming the physical realization of the GiBS, optimized geometry.

\subsection{Optical Characterization and Results}

\begin{figure}[h!]
    \centering
    \includegraphics[width=\textwidth]{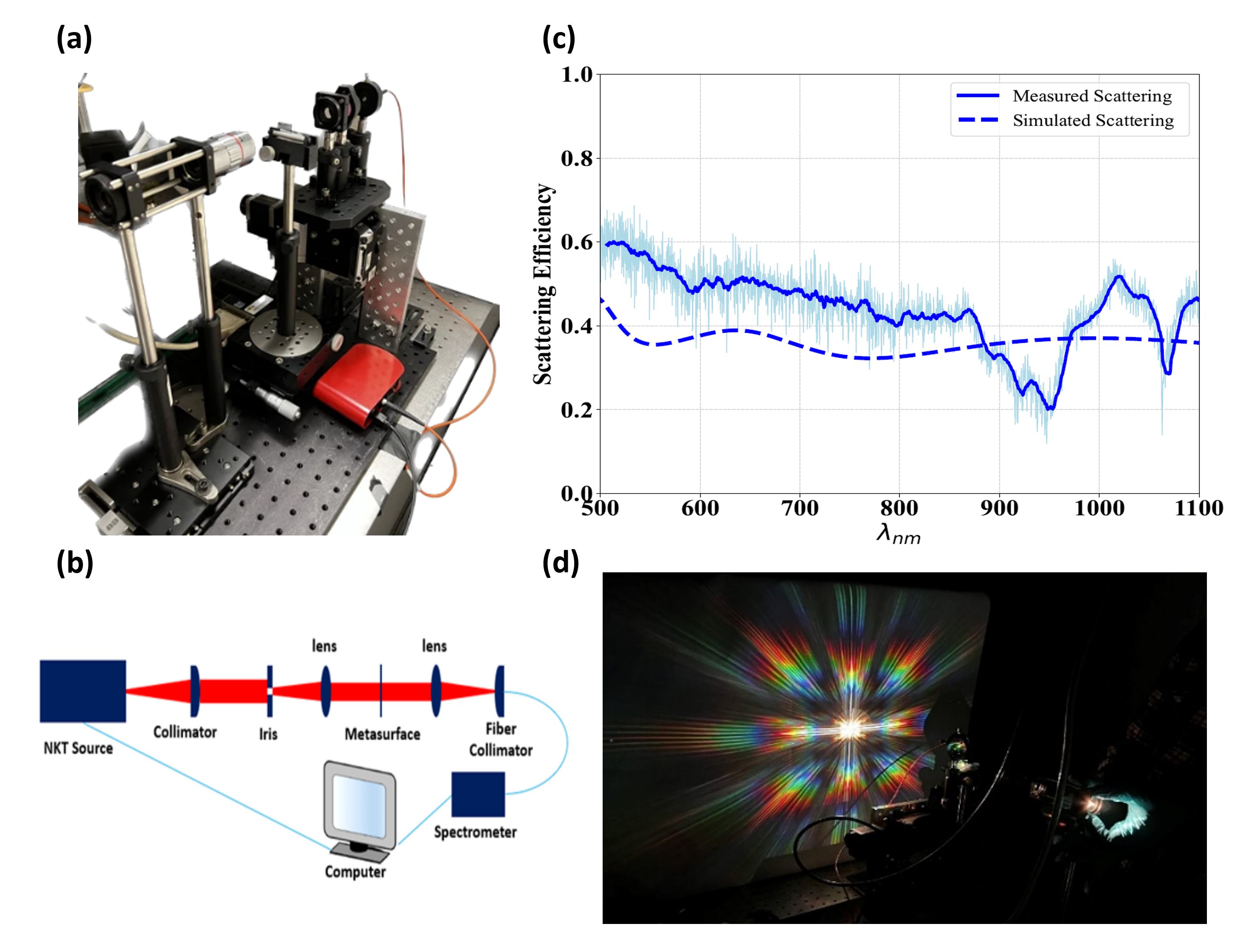}
    \caption{Optical characterization of the fabricated PEDOT:PSS metasurface. 
    (a) Photograph of the custom-built angle-resolved spectroscopy setup used for scattering measurements. 
    (b) Schematic of the optical configuration, showing the SuperK FIANIUM FIU-6 source, collimation optics, metasurface sample mounted on a motorized rotation stage, and the fiber-coupled Thorlabs CCS spectrometer on a Newport BGM120CC motorized goniometer enabling angular scanning from –45° to +45°. 
    (c) Comparison of measured (solid blue) and simulated (dashed blue) scattering efficiency spectra of the fabricated metasurface (Figure~\ref{fig:sem_pedot}) . 
    The measured data were post-processed using a moving-average filter to suppress high-frequency noise while preserving broadband features. 
    The slightly higher measured amplitude originates from diffuse and edge scattering effects inherent to the fabricated device.  
    (d) Photograph revealing strong angular scattering under broadband illumination of the same metasurface (Figure~\ref{fig:sem_pedot}), consistent with the simulated wide-angle response.}
    \label{fig:results}
\end{figure}

The optical performance of the fabricated PEDOT:PSS metasurface was characterized using a custom-built, angle-resolved spectroscopy setup. As shown in Figure~\ref{fig:results}a, the illumination was provided by a SuperK FIANIUM FIU-6 broadband source (NKT Photonics), delivering a high-stability, wide-spectrum optical signal from the visible to the near-infrared wavelength range. The output from the source was first spatially filtered and passed through an adjustable aperture, after which it was collimated to approximate a plane-wave excitation. The collimated beam was then directed onto the metasurface sample, which was mounted on a motorized rotation stage. By rotating this stage, the incidence angle on the metasurface could be precisely varied to probe different excitation conditions.

The scattered light was collected using an output arm positioned opposite to the input arm, equipped with a fiber-coupled spectrometer (Thorlabs CCS series) mounted on a Newport BGM120CC motorized goniometer, as shown in Figure~\ref{fig:results}a. This configuration enabled angular-resolved measurements of the scattered intensity from –45° to +45° relative to the surface normal. The output arm’s collection optics were aligned such that the scattered beam was efficiently coupled into the detection fiber and directed to the spectrometer for spectral analysis. The complete optical path, including the collimation optics, focusing objective, sample stage, and output goniometer, is schematically illustrated in Figure~\ref{fig:results}b. Each spectrum was recorded relative to a reference SiO$_2$ substrate under identical illumination conditions to remove the source spectral bias. Multiple scans were averaged, and a moving-average filter was applied to the experimental data to suppress high-frequency noise while retaining the overall spectral evolution. The processed experimental data are compared with full-wave simulations in Figure~\ref{fig:results}c, showing a wideband scattering response with good agreement with the theoretical response for the inverse-designed structure. 

Under broadband illumination, the metasurface exhibited vivid, multi-hued scattering across the visible and near-infrared regions. As shown in Figure~\ref{fig:results}d, white-light illumination produced strong angular dispersion and color separation, confirming the broadband, nonlocal optical behavior designed by the GiBS framework. Figure~\ref{fig:results}c shows that the measured scattering efficiency follows the simulated trend across the 500–-1100\,nm wavelength range, capturing the principal resonant features and broadband envelope predicted during design. The slightly higher experimental amplitude can be attributed to fabrication-realistic effects that are not captured in idealized full-wave simulations. First, the finite lateral size of the fabricated device (\(\sim250\times250~\mu\text{m}^2\)) introduces edge diffraction and near-field leakage, which redistribute optical power into the collection cone of the detector. Second, the 400\,nm-thick PEDOT:PSS nanopillars exhibit some sidewall roughness and minor rounding resulting from etching and solvent reflow, increasing diffuse scattering that adds to the measured intensity. Third, small refractive-index and thickness inhomogeneities in both the PEDOT:PSS layer and the underlying SiO\(_2\) substrate cause local phase deviations and scattering from interfacial index gradients. Collectively, these effects enhance the total measured signal without altering the overall spectral profile or resonance positions.

The strong spectral correspondence between simulation and experiment in Figure~\ref{fig:results}c confirms that the structures generated by the GiBS inverse-design framework accurately realize the targeted broadband response in practice. By coupling parametric basis representations with latent-space optimization and fabrication-aware constraints, GiBS enables the systematic discovery of large-scale, experimentally feasible metasurfaces that exhibit robust, wideband optical behavior. 

The metasurface design process here serves as an example of how GiBS enables large-scale metastructure optimization that would be prohibitively expensive for traditional TO methods. In a conventional TO workflow, optimizing a device on the scale of $30\lambda \times 30\lambda$ with subwavelength pixel resolution would require handling millions of design variables, leading to extreme memory and computational demands. Moreover, gradient-based TO algorithms are inherently best suited for \emph{many-to-few} mappings, where numerous design parameters target one or a few objective functions. In contrast, broadband metasurface design typically requires simultaneous optimization across hundreds of wavelengths, resulting in a \emph{many-to-many} optimization regime with highly correlated gradients that are difficult to converge. Such constraints make large-scale, broadband TO computationally challenging.
Similarly, dataset-driven inverse design approaches that rely on random geometric sampling struggle to cover the relevant optical response space, as shown in earlier sections.

\section{Conclusion}

Here, we introduced and experimentally validated the GiBS framework, an inverse-design strategy that integrates parametric basis modeling, representation learning, and fabrication-aware optimization for large-scale, nonlocal metasurfaces. By replacing discrete, pixel-based representations with continuous basis functions, GiBS compresses the high-dimensional nanophotonic design space into a compact and interpretable form that supports smooth, asymmetric, and aperiodic geometries essential for broadband responses. The latent-space learning module enables efficient exploration of structure–response correlations, while the fabrication-aware parameterization ensures robustness to process-induced variations. The framework was validated through the realization of a broadband scattering metasurface. The good agreement of the measured responses with theoretical predictions confirmed that GiBS-designed structures can faithfully translate inverse-designed geometries into experimentally realizable, high-performance optical components. The results shown here highlight GiBS as an effective AI-driven inverse-design route for producing physically consistent, multifunctional (e.g., broadband) fabrication-tolerant metasurfaces beyond the reach of traditional topology or data-driven optimization approaches. Its flexibility makes it extendable to multilayer and re-configurable systems, while future integration with gradient-informed optimization or active learning could further accelerate convergence and enable multifunctional metasurfaces.

\section*{Acknowledgments}
This research was developed with funding from the Defense Advanced Research Projects Agency (DARPA) under grant number FA8650-22-C-7207 (Coded Visibility). The views, opinions and/or findings expressed are those of the authors and should not be interpreted as representing the official views or policies of the Department of Defense or the U.S. Government. The authors gratefully acknowledge support from the Georgia Institute of Technology’s Institute for Matter and Systems (IMS).

% ---- References (SPIE style)
\bibliography{ref}
\bibliographystyle{spiejour}

\end{spacing}
\end{document}